\documentclass[twocolumn]{aastex63}
\usepackage[mediumspace,mediumqspace,Grey,squaren]{SIunits}

\graphicspath{{./}{figures/}}

\submitjournal{ApJ}

\shorttitle{Stellar populations in NGC 1978}
\shortauthors{Li et al.}

\begin{document}

\title{Multiple stellar populations at less evolved stages: detection of chemical variations among main-sequence dwarfs in NGC 1978}

\correspondingauthor{Chengyuan Li}
\email{lichengy5@mail.sysu.edu.cn}

\author{Chengyuan Li} 
\affil{School of Physics and Astronomy, Sun Yat-sen University, Daxue Road, Zhuhai, 519082, China}

\author{Baitian Tang} 
\affil{School of Physics and Astronomy, Sun Yat-sen University, Daxue Road, Zhuhai, 519082, China}

\author{Antonino P. Milone}
\affiliation{Dipartimento di Fisica e Astronomia ``Galileo Galilei'', Univ. di Padova, Vicolo dell'Osservatorio 3, Padova, IT-35122, Italy}
\affiliation{Istituto Nazionale di Astrofisica - Osservatorio Astronomico di
Padova, Vicolo dell'Osservatorio 5, Padova, IT-35122, Italy}

\author{Richard de Grijs}
\affiliation{Department of Physics and Astronomy, Macquarie
University, Daxue Road, Sydney, NSW 2109, Australia}
\affil{Centre for Astronomy, Astrophysics and Astrophotonics, 
Macquarie University, Daxue Road, Sydney, NSW 2109, Australia}
\affiliation{International Space Science Institute--Beijing, 1
Nanertiao, Zhongguancun, Hai Dian District, Beijing 100190, China}

\author{Jongsuk Hong}
\affiliation{Department of Astronomy, Yonsei University 50 Yonsei-Ro, Seodaemun-Gu, Seoul, Republic of Korea}
\affiliation{Korea Astronomy and Space Science Institute, Daejeon 34055, Republic of Korea}

\author{Yujiao Yang}
\affil{Department of Astronomy, School of Physics, Peking University, Yi He Yuan Lu 5, Haidian District, Beijing 100871, China}
\affiliation{Department of Physics and Astronomy, Macquarie University, Daxue Road, Sydney, NSW 2109, Australia}

\author{Yue Wang} 
\affiliation{Key Laboratory for Optical Astronomy, National
Astronomical Observatories, Chinese Academy of Sciences, 20A Datun
Road, Beijing 100101, China}

\begin{abstract}
Multiple stellar populations (MPs) with different chemical compositions are not exclusive 
features of old GCs (older than 10 Gyr). Indeed, recent studies reveal that younger 
clusters ($\sim$2--6 Gyr-old) in the Magellanic Clouds also exhibit star-to-star chemical variations 
among evolved stars. However, whether MPs are present among less evolved dwarfs of these  intermediate-age clusters is still unclear. In this work, we search for chemical variations among GK-type dwarfs in the $\sim$2 Gyr-old cluster NGC 1978, which is the youngest cluster with MPs.  We exploit deep ultraviolet and visual observations from the Hubble Space Telescope to constrain the  nitrogen (N) and oxygen (O) variations among MS stars. To do this, we compare appropriate photometric diagrams that are sensitive to N and O with synthetic diagrams of simple stellar populations 
and MPs. We conclude that the G- and K-type MS stars in NGC\,1978 host MPs. Our statistical analysis shows that the fraction of N-rich stars ranges from $\sim$40\% to $\sim$80\%, depending on the detailed distributions of nitrogen and oxygen. 
\end{abstract}


\section{Introduction} \label{S1}
Chemical variations among member stars of star clusters, which imply the presence 
of multiple stellar populations (MPs), are detected in intermediate-age ($\sim$2--6 
Gyr-old) Magellanic Cloud clusters \citep{Mart17a,Milo20a}. These findings expand 
the phenomenon of MPs to star clusters spanning a more extensive age range \citep{Mart18a,Li19a}. 
Like most old globular clusters (GCs), these intermediate-age clusters in the Magellanic Clouds 
exhibit clear signatures of chemical variations in He, C, N and O among their evolved stars 
\citep[e.g.,][]{Milo20a}. Once MPs appear in star clusters, seem to follow 
the same correlation between the number fractions and the host cluster masses, regardless 
of their age and the nature of their parent galaxy \citep{Milo20a}.

For Galactic GCs (GGCs), MPs are suggested to have a primordial origin. Intermediate-age 
AGB stars \citep{Derc10a,Dant16a}, fast-rotating massive main-sequence (MS) stars \citep{Decr07a}, 
massive interacting binaries \citep{deMi09a}, or just a single supermassive 
MS star \citep{Deni14a} may play a role as polluters of the second generation (SG) stars
\citep[and their combination][]{Wang20a}. MPs are known as a global feature 
for stellar populations at different evolutionary stages in GCs. Most works focus on MPs among the 
bright giant stars since they are the most luminous sources in most GCs \citep[e.g.,][]{Yong08a,Tang17a,Wang17a,Lee19a}. For GGCs, observations also provide substantial 
evidence which supports the presence of MPs among less-evolved stars, such as MS stars 
or sub-giant branch (SGB) stars \citep[e.g.,][]{Lard12a}. MPs are found 
among the bottom-MS stars ($\sim$0.15$M_{\odot}$) when photometry with deep exposures 
is involved \citep{Milo19a}. These findings rule out the possibility that MPs can form  
through standard stellar evolutionary processes such as evolutionary mixing \citep{Deni90a}. 

As discussed, understanding whether MPs have a primordial or an evolutionary origin is 
crucial to constrain their formation mechanisms.
For most Galactic open clusters (OCs), their photometry is strongly hampered by the large 
reddening caused by dust in the Galactic disk. This makes massive clusters in the 
Magellanic Clouds better targets. However, because of the large 
distance, whether the less evolved dwarfs in the Magellanic Clouds clusters could 
harbor MPs, is yet to be examined. It remains unclear if the MPs in Magellanic 
Cloud clusters could have the same origin as those in GGCs. Thanks to ultra-deep 
exposures in ultraviolet (UV) and optical passbands with the {\sl Hubble Space Telescope} 
({\sl HST}), some indirect evaluations of chemical spreads among MS stars in the Magellanic 
Clouds have become feasible. Based on UV-optical frames with deep exposures, the 
Small Magellanic Cloud (SMC) cluster NGC 419 ($\sim$1.4 Gyr-old) is proved to have no 
signature of MPs among its MS stars \citep{Cabr20a,Li20a}. But this does not lead 
to the conclusion that the origin of MPs in Magellanic Cloud clusters is different from that 
in GGCs, because NGC 419 does not exhibit any feature of MPs in its red-giant branch 
\citep[RGB,][]{Mart17a} either, making it a genuine SSP cluster. A similar study of the 
RGB of the cluster NGC 1783 was performed by \cite{Zhang18a}, which reported no 
evidence of chemical variations among these evolved giants. These clusters seem to 
define a minimum age limit for MP clusters \citep[$\sim$2 Gyr,][]{Mart18a}. Indeed, we 
have SSP GCs older than 2 Gyr but no MP GCs younger than 2 Gyr.

For clusters older than this age limit, NGC 1978 is one of the youngest clusters 
($\sim$2.3 Gyr) known to have MPs:  studies have reported the 
presence of internal chemical variations of C, N, O with a nitrogen variation of up 
to $\Delta{\rm [N/Fe]}$=0.5 dex \citep{Mart18a}, and a negligible helium 
spread \citep[$\Delta{Y}=0.002$ dex,][]{Milo20a}. All these conclusions were 
drawn based on analyses of RGB stars. It is, therefore, necessary to examine if its less 
evolved dwarfs could exhibit a similar feature of MPs. In this work, based on the 
depth of the observations, we used a similar method as in \cite{Li20a} to 
examine if NGC 1978 could exhibit a signature of MPs among its GK-type MS 
stars.  

In this article, the data reduction is discussed in Section 2. The methods we used and the main 
results are presented in Section 3. Our main conclusions are summarized in Section 4.

\section{Data Reduction} \label{S2}
Our datasets are derived from two observational programs observed through the 
{\sl HST}'s Ultraviolet and Visiable Channel (CENTER aperture) of the Wide Field 
Camera 3 (UVIS-CENTER/WFC3), they are programs with an ID of General Observer 
(GO-)14069 and GO-15630 (PI: N. Bastian). In Table \ref{T1}, we present the details of 
the observational frames. The passbands we used in this work include F275W, F343N, F438W 
and F814W, with total exposure times of 17,970 s, 3975 s, 2475 s and 2334 s, 
respectively. 

\begin{table*}
  \begin{center}
\caption{Description of the used observations.}\label{T1}
  \begin{tabular}{l | l l l l l l}\hline
    Rootname      &  Camera (Aperture)	  & Exposure time & Filter & Program ID \\\hline
    idxz12exq	& UVIS-CENTER/WFC3 & 1493.0 s	 & F275W	& GO-15630 & \\
    idxz12eyq	& UVIS-CENTER/WFC3 & 1498.0 s	 &  F275W & &  \\
    idxz12ezq	& UVIS-CENTER/WFC3 & 1500.0 s	 &  F275W & &  \\
    idxz12f1q	& UVIS-CENTER/WFC3 & 1499.0 s	 &  F275W & &  \\
    idxz12f2q	& UVIS-CENTER/WFC3 & 1501.0 s	 &  F275W & &  \\
    idxz12f4q	& UVIS-CENTER/WFC3 & 1502.0 s	 &  F275W & &  \\
    idxz13f6q	& UVIS-CENTER/WFC3 & 1493.0 s	 &  F275W & &  \\
    idxz13f7q	& UVIS-CENTER/WFC3 & 1495.0 s	 &  F275W & &  \\		
    idxz13f8q	& UVIS-CENTER/WFC3 & 1500.0 s	 &  F275W & &  \\
    idxz13faq	& UVIS-CENTER/WFC3 & 1498.0 s	 &  F275W & &  \\
    idxz13fbq	& UVIS-CENTER/WFC3 & 1492.0 s	 &  F275W & &  \\
    idxz13fdq	& UVIS-CENTER/WFC3 & 1499.0 s	 &  F275W & &  \\
    idxz14j0q	& UVIS-CENTER/WFC3 & 200.0 s	 &  F814W & &  \\
    idxz14j1q	& UVIS-CENTER/WFC3 & 349.0 s	 &  F814W & &  \\
    idxz14jrq	& UVIS-CENTER/WFC3 & 349.0 s	 &  F814W & &  \\
    idxz14jtq	& UVIS-CENTER/WFC3 & 200.0 s	 &  F814W & &  \\
    idxz14juq	& UVIS-CENTER/WFC3 & 688.0 s	 &  F814W & &  \\
    idxz14jwq	& UVIS-CENTER/WFC3 & 348.0 s	 &  F814W & &  \\
    idxz14jyq	& UVIS-CENTER/WFC3 & 200.0 s	 &  F814W & &  \\\hline	
    icz611r6q	& UVIS-CENTER/WFC3 & 500.0 s	 &  F343N & GO-14069 & \\
    icz611r7q	& UVIS-CENTER/WFC3 & 1000.0 s	 &  F343N & &  \\
    icz611r8q	& UVIS-CENTER/WFC3 & 800.0 s	 &  F343N & &  \\
    icz611raq	& UVIS-CENTER/WFC3 & 425.0 s	 &  F343N & &  \\
    icz611rcq	& UVIS-CENTER/WFC3 & 450.0 s	 &  F343N & &  \\
    icz611rdq	& UVIS-CENTER/WFC3 & 800.0 s	 &  F343N & &  \\
    icz611rfq	& UVIS-CENTER/WFC3 & 750.0 s	 &  F438W & &  \\
    icz611rhq	& UVIS-CENTER/WFC3 & 650.0 s	 &  F438W & &  \\
    icz611rjq	& UVIS-CENTER/WFC3 & 75.0 s	 &  F438W & &  \\
    icz611rlq	& UVIS-CENTER/WFC3 & 120.0 s	 &  F438W & &  \\
    icz611rmq	& UVIS-CENTER/WFC3 & 460.0 s	 &  F438W & &  \\
    icz611roq	& UVIS-CENTER/WFC3 & 420.0 s	 &  F438W & &  \\
    \hline
  \end{tabular} 
  \end{center} 
\end{table*} 

We perform point-spread-function (PSF) corrected photometry to the pixel-based 
charge transfer efficiency corrected frames (`{\tt \_flc}'), using the package {\sc Dolphot2.0} 
\citep{Dolp11a,Dolp11b,Dolp13a}. When adding fake stars, Poissonian noise is always 
applied to their PSFs. The standard data reduction procedures include 
bad pixel masking, splitting frames based on CCD chips and background correction. 
After that we perform the {\tt dolphot} command to frames belong to the same 
observational program. The {\tt dolphot} can automatically read the World Coordinate 
System (WCS) header information from each frame for alignment, enable a full stellar 
catalog including stellar magnitudes in different passbands. From the raw stellar catalog, 
we removed stars if they have any of these features: (1) Their stellar magnitude in any passband 
is 99.99 (flagged by the {\sc Dolphot} if these are failed measurement). (2) Their stellar types 
are not ``good star'' (flagged by the {\sc Dolphot} with a type-ID of 1). (3) They are likely 
cosmic rays or extended sources with a too low ($\leq$0.3) or too high ($\geq$0.3) sharpness. 
(4) Their crowding parameter is too high ($\geq$0.1 mag). Except for the crowding, all other 
parameters are decided followed by the optimal suggestions from the manual of 
{\sc Dolphot2.0} \citep{Dolp13a}. The crowding tells how much brighter the star would 
have been measured had nearby stars not been fit simultaneously. Because GK-type dwarfs 
are usually severely affected by crowding in a cluster, high crowding would hamper the 
reliability of our results. Under this adoption, we confirmed that even a maximum crowding (0.1 mag) 
would not exceed 20\% of the photometric uncertainty of stars for analysis.  
The average crowding parameter for our sample stars is only 0.02 mag. 

The spatial coordinate of stars in the raw stellar sample is in X \& Y, which describes their 
2D positions in the CCD. We transferred their CCD positions to the right ascension ($\alpha_{\rm J2000}$) and declination ($\delta_{\rm J2000}$) using the WCS matrix identified from the header of the reference frames. 

At this stage, we will obtain two stellar catalogs, corresponding to two different observational programs GO-14069 and GO-15630. These two stellar catalogs describe the same cluster NGC 1978 in 
different passbands (see Table \ref{T1}). We combined these two stellar catalogs through cross-matching their spatial 
coordinates ($\alpha_{\rm J2000}$ and $\delta_{\rm J2000}$). The detection limit (the magnitude where the stellar completeness is 
$\sim$50\%) is determined by the combination of F275W, F343N, F438W and F814W passbands observation. 
That means if a star is not detected in any one of these passbands, we treat it as failed detection and will contribute 
to the incompleteness. In this work, in order to improve the reliability, we only select stars that are 0.3 mag brighter 
than the detection limit (GK-type stars). The completeness for our stars of interest ranges from 65.1\% to 93.4\%, 
depending on the artificial star (AS) test (introduced below). The average completeness for these stars is greater than 80\%.   



\begin{figure*}[!ht]
\includegraphics[width=2\columnwidth]{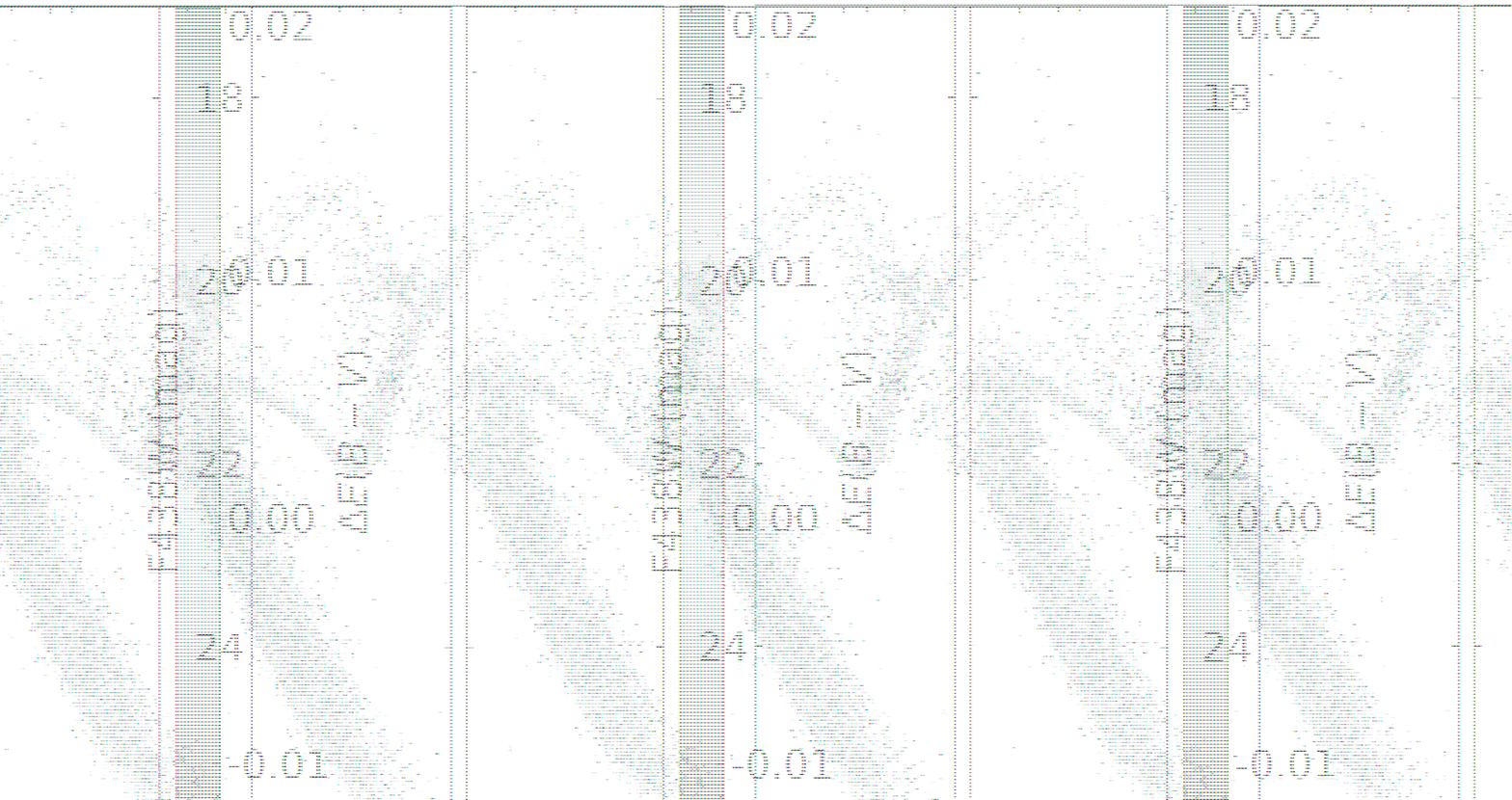}
\caption{Raw CMD (left) and reddening variation corrected CMD (right) for NGC 1978, for each star, 
color indicate their reddening difference to the average value ($\delta{E(B-V)}$, mag).}
\label{F1}
\end{figure*}

We first correct the differential reddening, the reddening variation of stars across the entire cluster region, 
for our sample stars using the same method as \cite{Milo12a}. Their method cannot determine the 
absolute reddening for individual stars. Instead, it would correct stars at different positions to 
the median reddening value of the whole stellar population. We examined the 
distribution of differential reddening for stars in NGC 1978 using their (F438W $-$ F814W 
vs. F438W) color-magnitude diagram (CMD). In Fig.\ref{F1} we show this 
CMD before/after correcting for differential reddening. We report an average differential 
reddening degree of $\delta{E(B-V)}$=0.006 mag, with a maximum value of  
$\delta{E(B-V)}$=0.021 mag. From Fig.\ref{F1} (left panel) we can see that before we 
correct for differential reddening, stars with large differential reddening show a 
preferential distribution around the red side of the MS. In the right panel 
of Fig.\ref{F1} we confirmed that the differential reddening corrected CMD does not 
show this reddening-color correlation any more.  

As introduced by \cite{Milo12a}, in addition to the differential reddening, some unmodelable 
PSF variations may add additional noise to the observation, leading to color variations for different 
color baselines. This effect could be very significant for faint stars with low signal to noise ratios. 
We followed the same procedures as \cite{Milo12a} to correct for the 
variations in F275W $-$ F343N, F343N $-$ F438W, because the color index we used is the 
combination of these two colors (See Section 3). However, note that even though we have 
corrected for possible variations in both the directions of reddening and different colors, some 
statistical residual would still remains. 


The field-of-view (FoV) of our observations is very small. For the combined stellar catalog, 
the maximum angular radius from the cluster center is only $\sim$2.1 arcmin, or $\sim$30.2 parsec (pc), 
which is smaller than the angular size of NGC 1978 \citep[2.7--4.0 arcmin, or 38.9--57.6 pc][]{Bona10a}. 
As a result, we are not able to find an appropriate reference field for comparison to estimate the background 
contamination level. As a second-best choice, we only select stars that are located within 29.2 arcsec (7.0 pc) 
\citep[the projected half-light radius, $R_{\rm hl}$,][]{Milo20a} for our analysis as in this inner region most stars 
should be cluster members. We select a sample of stars beyond four times the half-light radius ($R\geq$116.8 arcsec, 28.0 pc) 
as reference field to reduce the possible field contamination. The field decontamination consisted of two steps: 
we divided the observed color index of both the field and the cluster into several bins (with a bin size of 0.1 mag). Then we 
randomly subtract the corresponding number of stars in the cluster from the same color index bin (with 
completeness and area difference corrected). We emphasize that the radius we used for selecting field stars 
is smaller than the size of NGC 1978. Therefore we may have overestimated the field contamination level. 

Finally, the stellar sample we used is the combined stellar catalog with differential reddening, color variations 
and field contamination statistically corrected. It includes stars with reliable detections and low 
sharpness, crowding, in F275W, F343N, F438W and F814W passbands.  



\section{Methods and Results}
\subsection{Photometric and spectroscopic models}
In this work, we used the color index
\begin{widetext}
\begin{equation}
C_{\rm F275W,F343N,F438W} = {\rm (F275W-F343N)-(F343N-F438W)}
\end{equation}
\end{widetext}
to reveal the possible presence of MPs along the MS of NGC 1978. For 
late-type stars, their UV band spectrum ($\leq$300 nm) contains a lot of OH absorption 
bands. In the meantime, the NH and CH absorption bands are centered around 
$\sim337$ nm and $\sim430$ nm, respectively. For GGCs with MPs, a typical feature 
is the N enriched population stars would be depleted in O and C. \cite{Zenn19a} proved 
that $C_{\rm F275W,F343N,F438W}$ can maximize the separation between the primordial 
and enriched stellar populations. 

We applied the MESA Isochrone and Stellar Tracks \citep[MIST;][]{Paxt11a,Paxt13a,Paxt15a,Choi16a,Dott16a} 
to find the best fitting model to the observation. We determined the best fitting global parameters of 
$\log{(t/{\rm yr})}$=9.38$\pm$0.02 ($\sim$2.4 Gyr), [Fe/H]=$-$0.55$\pm$0.05 dex, 
 $A_{V}$=0.20$\pm$0.01 mag and $(m-M)_0$=18.50$\pm$0.05 mag. 
These parameters are close to those of \cite{Mart18a}. NGC 1978 does 
not exhibit an extended MS turnoff region, which is a feature produced 
by fast stellar rotation \citep[i.e.,][]{Li14b,Cord18a}, the adopted rotational 
velocity for the best fitting isochrone is thus zero. The uncertainty associated with 
each parameter is the width of the generated grid for fitting. Here the best fitting 
model means they represent the optimal fitting to all four passbands. 

\begin{figure*}[!ht]
\includegraphics[width=2\columnwidth]{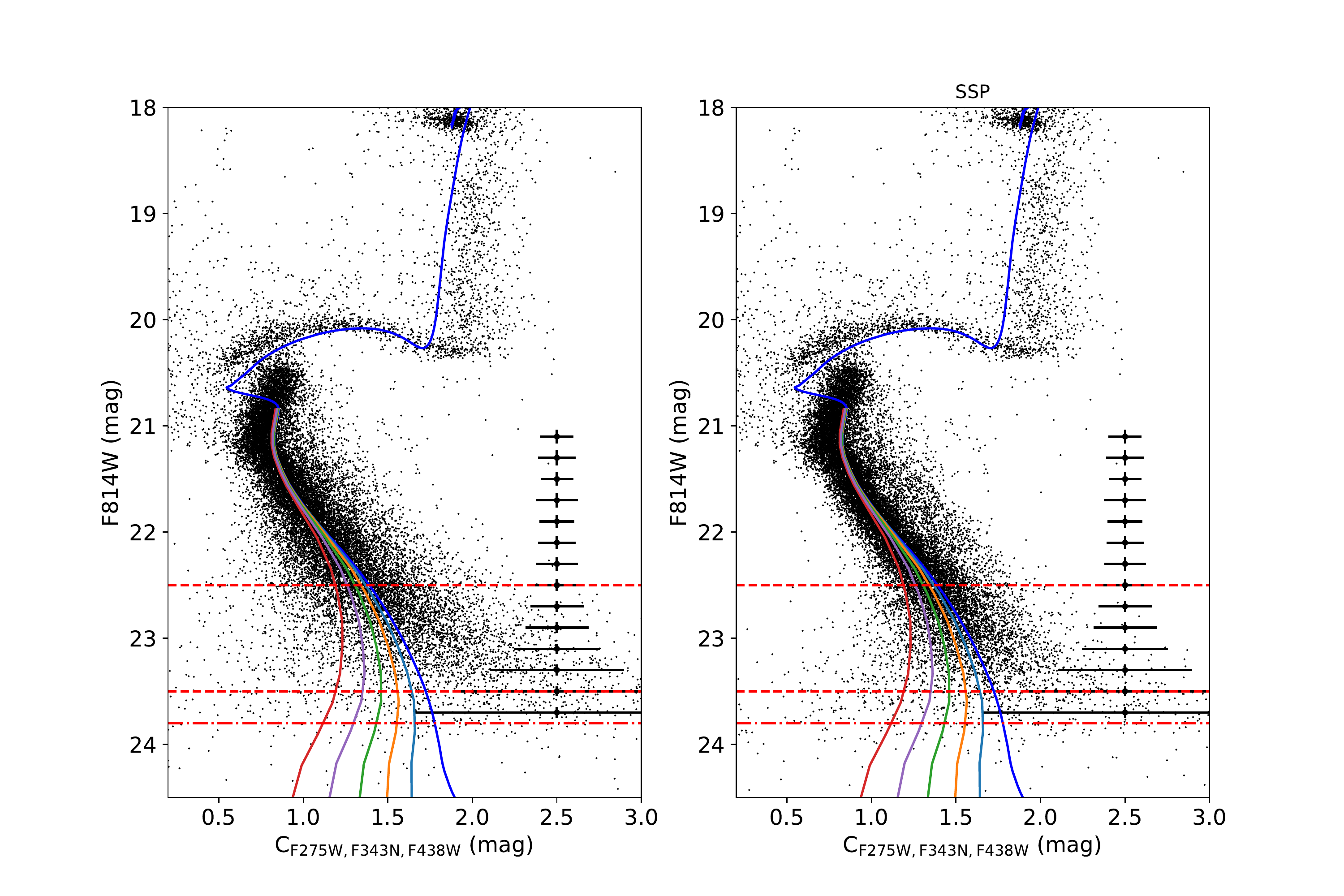}
\caption{$C_{\rm F275W,F343N,F438W}$ vs. F814W diagram of NGC 1978 (left panel). The blue 
curve is the best fitting isochrone. For the MS part, from right to left, curves located to the left 
side of the isochrone are loci for populations with $\Delta$[N/Fe]=0.2, 0.4, 0.6, 0.8 and 1.0 dex.
We only analyze stars located between the red dashed lines (22.5 mag$<$F814W$<$23.5 mag). 
The red dash-dotted line indicates the detection limit (completeness of 50\%, where F814W=23.8 mag). 
The red side of the MS spread can be explained by a combination of photometric uncertainties, 
unresolved binaries as well as contamination from field stars, as illustrated by the simulated 
diagram (right panel). In each panel, errorbars are attached in the right side (only for the MS part).}
\label{F2}
\end{figure*}

In this work, the isochrone is only used for generating enriched stellar models. 
To examine the effect of chemical enrichment for different passbands, we first 
generate a series of synthetic spectra using the iSpec \citep{Blan14a,Blan19a}. 
The radiative transfer code and line lists come from SPECTRUM\footnote{http://www.appstate.edu/$\sim$grayro/spectrum/spectrum.html}. MARCS model atmospheres and solar abundances were 
adopted \citep{Gust08a,Aspl09a}. For stars on the best-fitting 
isochrone, we calculated their chemically enriched counterparts. These 
enriched stellar spectra have the same global parameters ($\log g$, $\log{T_{\rm eff}}$, 
$X,Y,Z$) as the reference stars. But they are enriched in N and depleted 
in C and O, while the total abundance of the CNO remains constant. For each 
star, their chemically enriched counterparts will exhibit different feature in the F275W, 
F343N, F438W passbands (their difference in the F814W passband is very small, less 
than 0.001 mag), leading to measurable differences in photometry. For each star pair, we convolved 
both the ordinary and enriched stellar spectra with the different passbands and calculated 
their flux ratio. This ratio was converted into magnitude differences in different passbands. 
Stars with the same degree of chemical enrichment thus populate a locus with different 
$C_{\rm F275W,F343N,F438W}$. We calculated loci for $\Delta$[N/Fe]=0.2, 0.4, 0.6, 0.8 and 1.0 dex, 
and $\Delta$[O/Fe]=$\Delta$[C/Fe] = $-$0.02, $-$0.06, $-$0.13, $-$0.26 and $-$0.63 dex. 
As a toy model, we assumed that both oxygen and carbon will be depleted to  
the same degree. These combinations are determined by the constraint of $\Delta$[C+N+O/Fe]=0. 
Because MPs among red-giant stars in NGC 1978 have been well studied 
by \cite{Mart18a} and \cite{Milo20a}, we only calculated the MS part for each locus. In Fig.\ref{F2}, 
we present the $C_{\rm F275W,F343N,F438W}$ vs. F814W diagram of NGC 1978 (we 
selected F814W as the vertical axis because in this diagram the RGB and the MS are well 
separated), where we have also included the theoretical loci for the best fitting isochrone 
and enriched stellar populations. 

The model tells us that the chemically enriched stellar population would have a 
smaller $C_{\rm F275W,F343N,F438W}$ index than normal stars. This can be readily 
expected, because an N-enriched star would have deeper NH absorption at 337 nm, 
leading to a lower flux in the F343N passband, thus a larger magnitude. In the meantime, 
the N-enriched star would be O depleted. Therefore the OH dominated passband, 
F275W, would have a smaller magnitude. As a result, we obtained a smaller F275W$-$F343N 
color. In addition, the depleted C would produce shallower CH absorption at 430 nm, 
leading to a smaller F438W magnitude, and thus a larger F343N$-$F438W color. 
Their difference, the $C_{\rm F275W,F343N,F438W}$ index, is thus smaller than that for 
its ordinary counterpart. In Fig.\ref{F2} we see that the differences between different loci and the standard isochrone 
become more significant with increasing magnitude from F814W$\gtrsim$22 mag. 
However, with the magnitude increase, the photometric uncertainty increases as well. To 
find a balance between these two effects, we finally determine an optimal magnitude 
range of 22.5$>$F814W$>$23.5 mag (Fig.\ref{F2}). This magnitude range would 
select the sample stars with surface temperature between 5480 K and 6370 K 
(K- to late G-type). 

In the left panel of Fig.\ref{F2}, we present the observed $C_{\rm F275W,F343N,F438W}$ 
vs. F814W diagram of NGC 1978.  An asymmetry spread of $C_{\rm F275W,F343N,F438W}$ 
of the MS can be seen in this diagram. This is predominantly caused by the difference of photometric 
uncertainties between different passbands. The `red' side (with high $C_{\rm F275W,F343N,F438W}$) 
of the MS can be explained by photometric uncertainties and unresolved binaries. In the right panel 
of Fig.\ref{F2}, we present the $C_{\rm F275W,F343N,F438W}$ vs. F814W diagram of a simulated 
SSP sample. 

We used the same method as in \cite{Li20a} to mimic a real observation. We 
generated a sample of ASs following the standard isochrone and 
N-enriched loci with a Kroupa mass function \citep{Krou01a}. The total number of stars in 
each model population is 367,000. Unresolved binaries may play an additional role in the 
broadening of the MS. However, no studies have reported the MS-MS binary ratio 
for NGC 1978. We therefore used the same method as designed in \cite{Milo12a} 
to estimate the MS-MS binary fraction. If we assume a flat mass-ratio distribution, the total GK-type 
MS-MS binary fraction should be 50\%. However, we highlight that this must be a maximum 
binary fraction, as soft binaries (low mass and low mass-ratio binaries) would be quickly 
dissolved through three-body encounters in the denset region of the cluster \citep[e.g.,][]{Gell13a}. 
For the GK-type MS, both photometric uncertainties and unresolved binaries will lead 
to a broadening of the MS to the positive side of the $C_{\rm F275W,F343N,F438W}$ index. But the 
photometric uncertainty is much more important than binaries. Therefore, a variation in the  
binary fraction from 20\% to 50\% would not strongly affect our results. 

We generated six artificial stellar populations, including a standard SSP with 
$\Delta$[N/Fe]=0.0 dex, and a series of SSPs with $\Delta$[N/Fe]=0.2, 0.4, 0.6, 0.8 
and 1.0 dex. For each population, we divided them into 3,670 sub-samples, each 
containing 100 stars. These sub-samples were added to the raw images and recovered 
using {\sc Dolphot2.0} following the same adoption as for the real observation. That means we 
have repeated these procedures 7,340 times for each population (3,670$\times$2, because 
we have used two observational channels). Because both the real and the synthetic 
population stars are measured from the same frame using the same photometric procedure, 
they should suffer from the same photometric effects, such as photometric uncertainty, contaminations 
by real objects (cosmic rays, extended sources) or bad pixels. For ASs, 
we reduced them following the same criteria as for the real observation. 
The initial spatial distribution of the ASs was uniform. Because ASs are less affected 
by crowding, their average photometric uncertainty is smaller than real stars. 
We confirmed that once we have filtered all ASs by their crowding as for the 
real observation, the difference between their photometric uncertainties becomes 
negligible.
As introduced, the color variations in different baselines cannot be fully reduced 
from the observation due to the low signal to noise ratios of our sample stars. 
Because all CNO related molecules would be destroyed in hot stars. Therefore, the 
chemical variation would not affect early-type MS stars.
We decided to add additional color variations ($\sim$0.02--0.03 mag) to the ASs 
to mimic the observed width of the MS with F814W$\leq$21.5 mag. 

We then constructed five synthetic MPs with different N-enrichment (and C,O depletion) 
through mixing different SSPs. The fraction of each population stars with different 
chemical enrichment depends on the distribution of the chemical spread. We assumed 
two different chemical spread distributions: flat distribution and bimodal 
distribution (see next section). 

\subsection{Statistical Analysis}
\begin{figure*}[!ht]
\includegraphics[width=2\columnwidth]{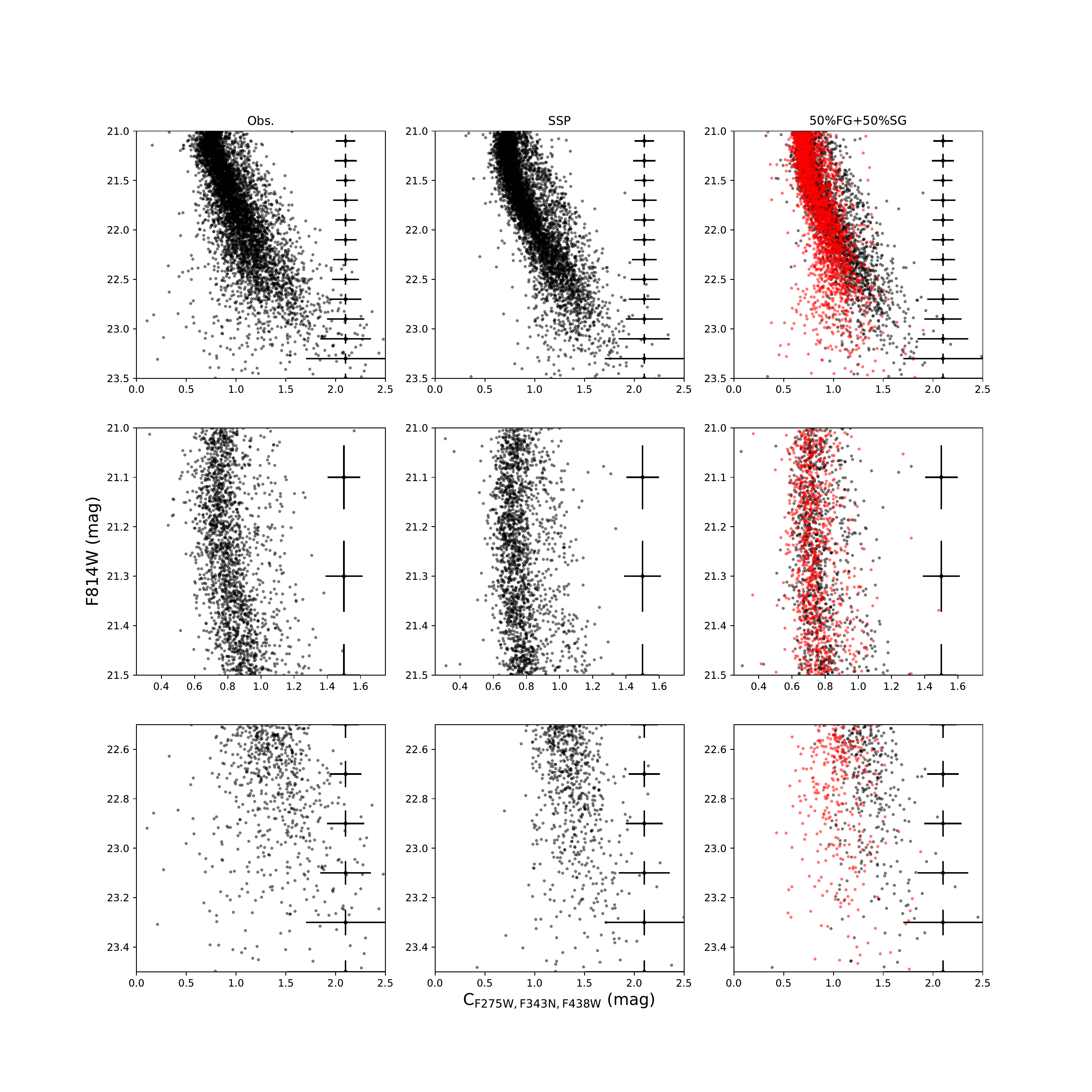}
\caption{Comparisons between the $C_{\rm F275W,F343N,F438W}$ vs. F814W diagrams of 
the observation (Left), the synthetic SSP (Middle) and the synthetic MPs where 50\% stars are 
enriched by $\Delta$[N/Fe]=1.0 dex (right, red circles). From top to bottom are diagrams for  
different magnitude ranges. (SG: second-generation, the chemically enriched population; FG: 
first-generation, the pristine population.)}
\label{F3}
\end{figure*}

We first undertake a visual inspection of the observation and the synthetic SSP 
and MPs.  The left, middle and right panels of fig. \ref{F3} are the $C_{\rm F275W,F343N,F438W}$ 
vs. F814W diagrams of the observation, the synthetic SSP and the synthetic MPs with 
50\% SG stars ($\Delta$[N/Fe]=1.0 dex), respectively. 
From top to bottom are diagrams for different F814W magnitude ranges. An overall impression 
of Fig.\ref{F3} is that the observed MS is indeed `fatter' than the synthetic SSP. 
In the bottom-left and middle panels we can see an 
excess of stars with $C_{\rm F275W,F343N,F438W}<1.0$ mag compare to the simulated SSP model. 
When consider a presence of stars with $\Delta$[N/Fe]=1.0 dex (red dots in the bottom-right panel), 
the difference is released.
This difference is negligible for the MS part in the range of 21.0$\leq$F814W$\leq$21.5 mag, which 
proves that we have correctly reproduced the spread where the effect of MPs is 
not important. For the MS part for F814W$>$21.5 mag, the difference becomes 
apparent, as illustrated in the bottom panels. As one can see, the observation exhibits an 
apparent excess of stars on the ‘blue’ side of the MS. This indicates a fraction of enriched 
population stars to reproduce the observation. The results presented in Fig.\ref{F3} are 
in contrast to NGC 419, which exhibits a high similarity between its observed MS and the 
synthetic SSP \citep[][their Fig.2]{Li20a}. 

\begin{figure*}[!ht]
\includegraphics[width=2\columnwidth]{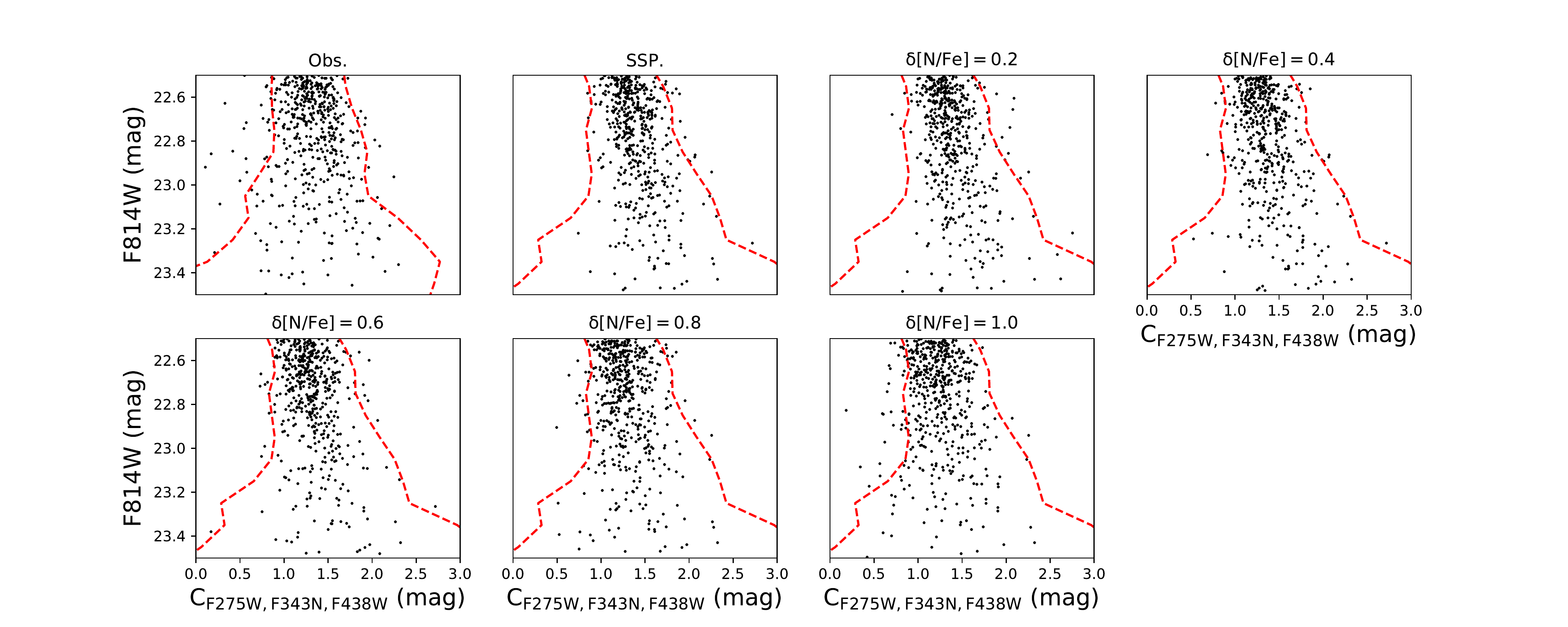}
\caption{$C_{\rm F275W,F343N,F438W}$ vs. F814W diagrams of the observation, the synthetic SSP, as well as MPs with different degrees of internal chemical spread. The red dashed lines are boundaries corresponding to $\pm$3 times the 
measurement uncertainties.
}
\label{F4}
\end{figure*}

We now discuss the effect of the measurement uncertainties, including photometric uncertainties as 
well as blending and PSF variations. These effects, in principle, should be reflected by ASs. In 
Fig.\ref{F4}, we present the observed  $C_{\rm F275W,F343N,F438W}$ vs. F814W diagram as 
well as two boundaries determined by tripling the measurement uncertainties and the ridge-line 
(top left panel). If assuming that the measurement uncertainty is Poisson-like, only a fraction of 0.3\% 
stars should be located outside of these two boundaries. However, in total we detected 2.0\% stars 
that are located in the right-side of the right boundary and 4.7\% stars that are located in the left-side 
of the left boundary. This is more than twenty times the expectation 
(2.0\%+4.7\% vs. 0.3\%). 

The excess of stars located in the right-side of the boundary can be explained by unresolved binaries 
and field contamination residuals. Using ASs, with 50\% unresolved binaries, the average fraction 
of stars in the right-side of the boundary is 1.7\% (derived from the synthetic SSP model). However, 
no other effect except for the measurement uncertainty and chemical enrichment, could contribute 
stars located in the left-side of the boundary. Indeed, using synthetic SSP, we only detected 0.5\% 
stars beyond the left boundary. The observed excess of stars in the `blue' side of the MS (blue-excess stars) 
is, therefore, have to be explained by a fraction of enriched population stars. 

\begin{figure}[!ht]
\includegraphics[width=\columnwidth]{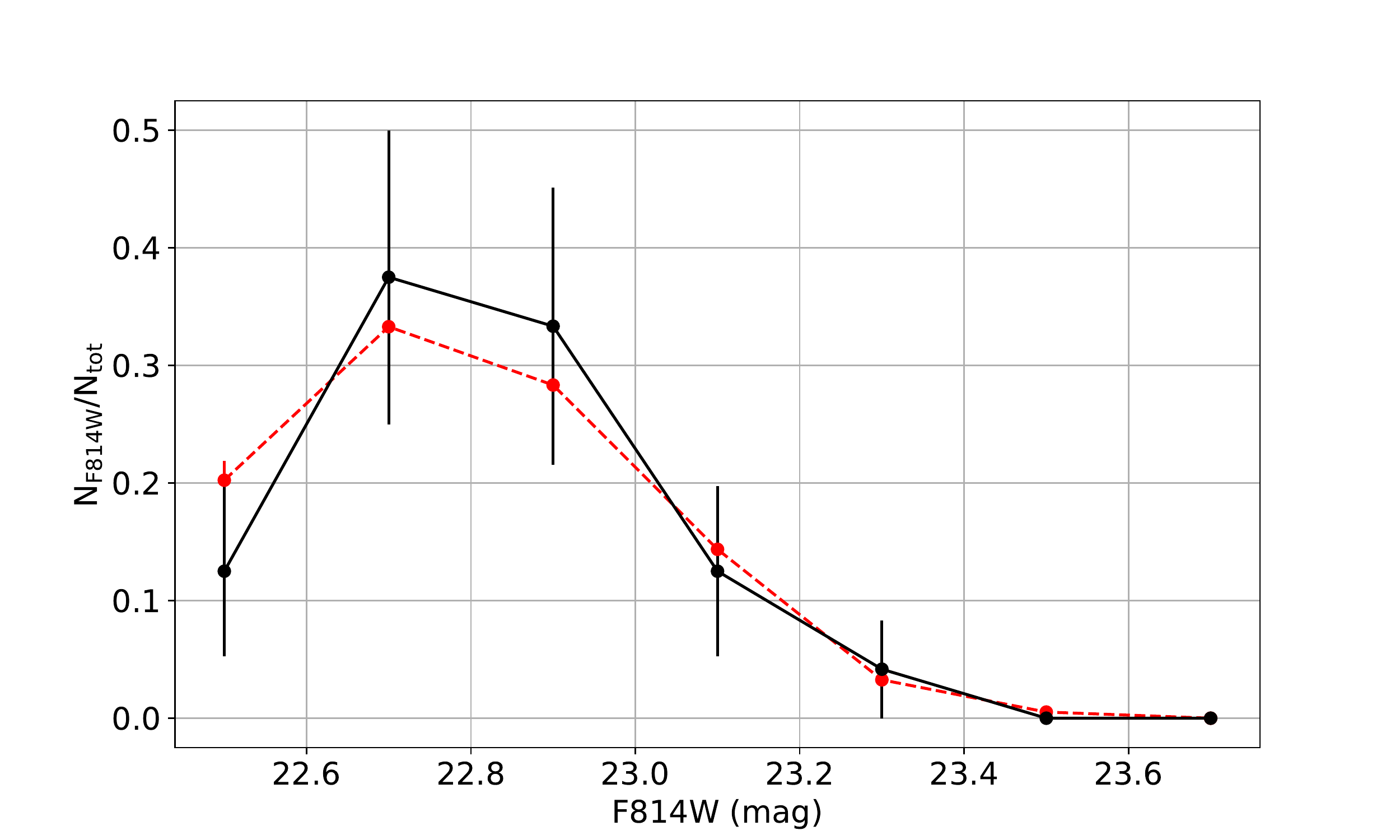}
\caption{Luminosity functions of blue-excess stars of the observation (black solid line) and the synthetic 
MPs (red dashed line) with an internal chemical spread of $\delta$[N/Fe]=1.0 dex. 
}
\label{F5}
\end{figure}

The first thing we explored is the degree of internal chemical spread. We generated five synthetic MPs 
composed of 50\% pristine population stars and 50\% enriched population stars with 
$\delta$[N/Fe]=0.2, 0.4, 0.6, 0.8, 1.0 dex (i.e., a bimodal distribution of chemical spread). As expected, 
the MS becomes wider in $C_{\rm F275W,F343N,F438W}$ with the increasing internal chemical spread. 
The average number fractions of blue-excess stars are 
0.8\%, 1.1\%, 1.9\%, 3.5\% and 8.2\%, respectively. In Fig.\ref{F4} we present one example of 
their $C_{\rm F275W,F343N,F438W}$ vs. F814W diagrams. The average number fractions are calculated 
through twenty runs of AS simulations. The luminosity function (LF) of blue-excess stars 
is not a random distribution. Most stars have their magnitude of F814W$\sim$22.7--22.9 mag, 
because in this magnitude range the photometric uncertainty is not large enough to mask the 
effect of chemical enrichment. We confirmed that the observed LF of these blue-excess stars 
is consistent with MPs with a maximum chemical spread of $\Delta$[N/Fe]=1.0 dex (Fig.\ref{F5}). 

 
\begin{figure*}[!ht]
\includegraphics[width=2\columnwidth]{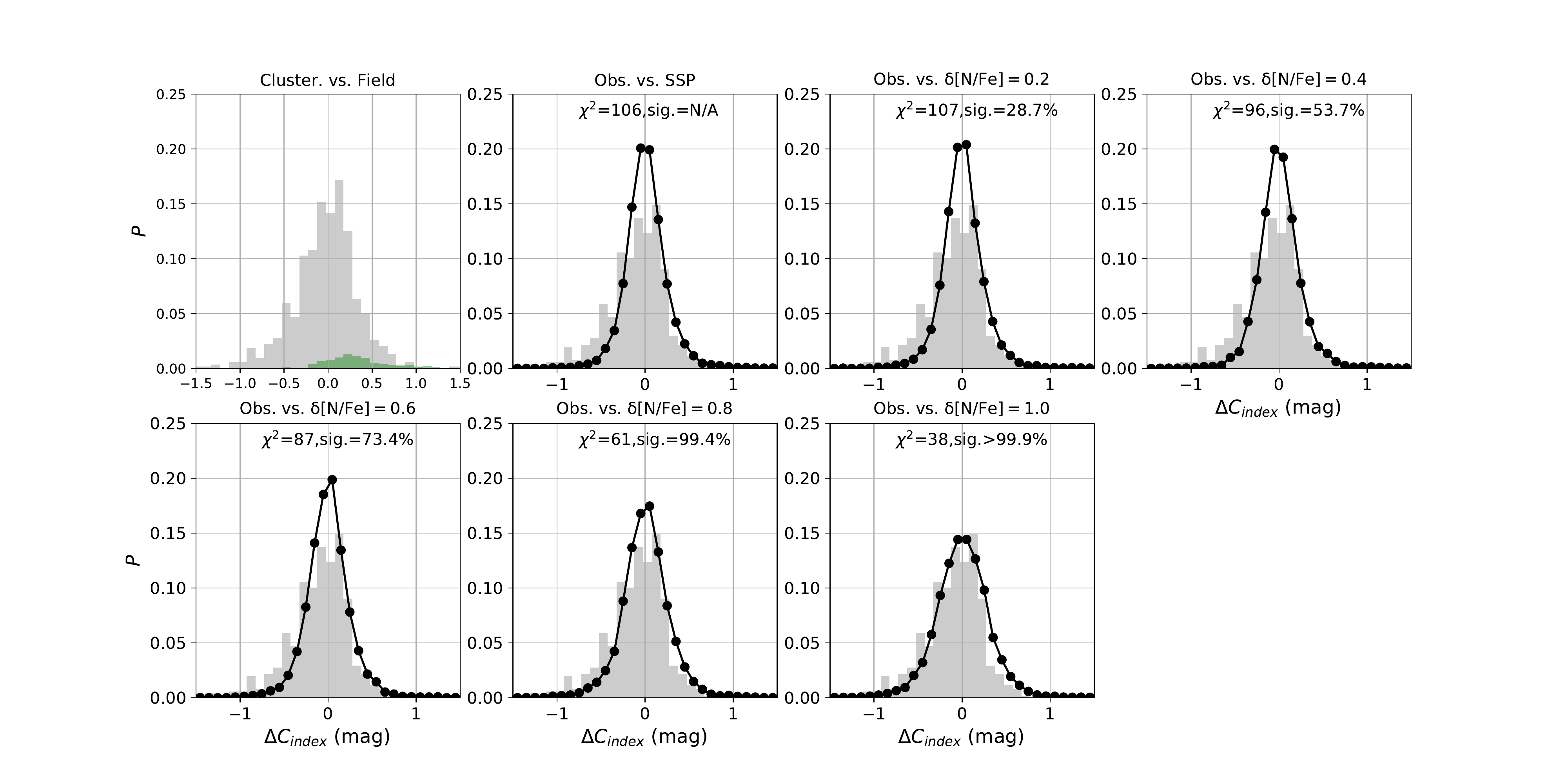}
\caption{Top left: normalized $\Delta{C_{\rm F275W,F343N,F438W}}$ distributions of both the cluster (grey) 
and `field' samples (green). The other panels are the $\Delta{C_{\rm F275W,F343N,F438W}}$ distribution of the field subtracted sample (grey histograms) versus that for the synthetic populations (black curves). The $\chi^2$ and the significance are indicated. 
}
\label{F6}
\end{figure*}

We also used an alternative method to quantify the similarity between the observation and the synthetic populations. 
We compared the distribution of $\Delta{C_{\rm F275W,F343N,F438W}}$ for the observed and artificial MS stars. 
For both the observation and synthetic populations, $\Delta{C_{\rm F275W,F343N,F438W}}$ is 
the deviation of the detected ${C_{\rm F275W,F343N,F438W}}$ for both the real and ASs to the MS ridge-line. 
The MS ridge-line is determined by connecting the median ${C_{\rm F275W,F343N,F438W}}$ at different 
F814W magnitudes in steps of 0.1 mag. We then calculate the similarity between the $\Delta{C_{\rm F275W,F343N,F438W}}$ distribution of the observation and synthetic populations, through a $\chi^2$ 
minimization method: 

\begin{equation}
\chi^2=\sum\limits_{n=1}^N\frac{(N_{\rm i}-N'_{\rm i})^2}{\sigma_{\rm i}}
\end{equation}

where $N$ is the number of observed stars within a given range centered at 
$\Delta{C_{\rm F275W,F343N,F438W}}$. $N_{\rm i}$ and $N'_{\rm i}$ are numbers of 
observed/simulated stars located in the $i$-th $\Delta{C_{\rm F275W,F343N,F438W}}$ range, 
respectively. $\sigma_{\rm i}$=$\sqrt{N_{\rm i}}$ is assumed a Poisson-like uncertainty for 
the number of stars in each bin. To improve the statistical significance, we only count bins 
containing at least 10 stars, and all other poorly populated bins were grouped 
together for the calculation of $\chi^2$. The $\chi^2$ indicates the similarity between the observation 
and the simulated populations. In principle, the smaller the $\chi^2$, the higher the similarity. 
In Fig.\ref{F6} we present the normalized $\Delta{C_{\rm F275W,F343N,F438W}}$ distribution of the 
observation (with field contamination statistically subtracted) versus that for the synthetic 
stellar populations with different $\delta$[N/Fe]. To reduce the noise, we have generated twenty synthetic 
populations with different chemical spreads and calculated their average $\chi^2$ values as 
representations. To estimate the significance of their $\chi^2$ difference, we repeated the $\chi^2$ 
calculation for the SSP model 1000 times. Then we counted how many times the $\chi^2$ of the SSP model 
would be lower than the MP model. As an example, in Fig. \ref{F6} we show that the $\chi^2$ of the MP 
model with $\delta$[N/Fe]=0.8 dex is 61, which is smaller than 996 of those 1000 $\chi^2$ values we 
calculated for the SSP model. We thus define its significance as 99.4\%.
Our result shows that the $\chi^2$ continues to decrease for an increasing 
$\delta$[N/Fe]. For MPs with $\delta$[N/Fe]=0.8 dex and 1.0 dex, 
their significances of $\chi^2$ are higher than 99\%. We thus suggest that this result could indicate
 the presence of a chemical spread, with $\delta$[N/Fe]$\geq$0.8 dex. 



We then explore the number fraction of enriched population stars. This depends on the 
distribution of the chemical spread. The real distributions of the chemical enrichment varies 
cluster to cluster. However, because of the large photometric uncertainties of the MS stars, 
we cannot resolve the detailed distribution of their chemical enrichment distribution. 

We only explored two extreme cases of chemical spread. The first model 
invokes a flat distribution of chemical spread. We set the pristine population as the 
dominant population, and all other populations with different $\Delta$[N/Fe] are 
homogeneously mixed. All synthetic MPs have an internal chemical 
spread of $\delta$[N/Fe]=1.0 dex. For example, a MPs with 50\% enriched population 
stars will contain 50\% pristine stars, and populations with $\Delta$[N/Fe]=0.2, 0.4, 0.6, 0.8, 
1.0 dex occupy 10\% each. We have generated nine MPs with enriched population stars 
ranging from 10\% to 90\%, calculated their $\Delta{C_{\rm F275W,F343N,F438W}}$ 
distributions, and then quantified their similarities to the observation through the $\chi^2$ 
minimization method introduced above.

\begin{figure*}[!ht]
\includegraphics[width=2\columnwidth]{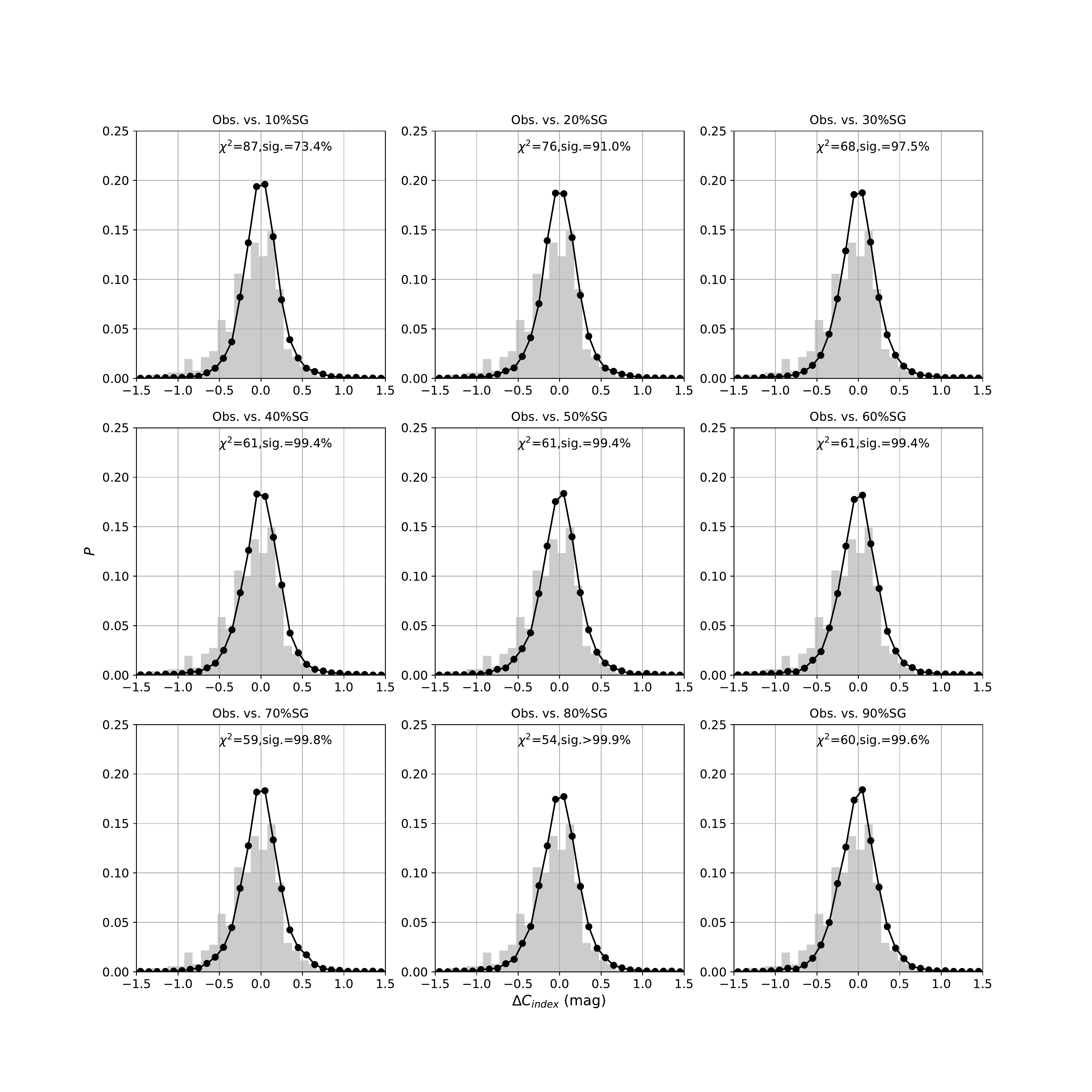}
\caption{$\Delta{C_{\rm F275W,F343N,F438W}}$ distributions of the observation (grey histograms) versus 
the synthetic MPs (black curves). The chemical distribution is flat.
}
\label{F7}
\end{figure*}

In Fig.\ref{F7} we present the $\Delta{C_{\rm F275W,F343N,F438W}}$ distributions. 
We find that the reproduction of the synthetic MPs is unsatisfactory. The optimal 
reproduction requires a very high fraction of enriched population stars (80\%). 
Through visual inspection, we find that even the model with 80\% enriched population 
stars would have a sharper $\Delta{C_{\rm F275W,F343N,F438W}}$ distribution than the 
observation, which may indicate that only the population with a high chemical 
enrichment (i.e.,$\delta$[N/Fe]=1.0 dex) would produce significant broadening of  
the MS. This forced us to find another probability of chemical spread, the bimodal 
distribution. 

\begin{figure*}[!ht]
\includegraphics[width=2\columnwidth]{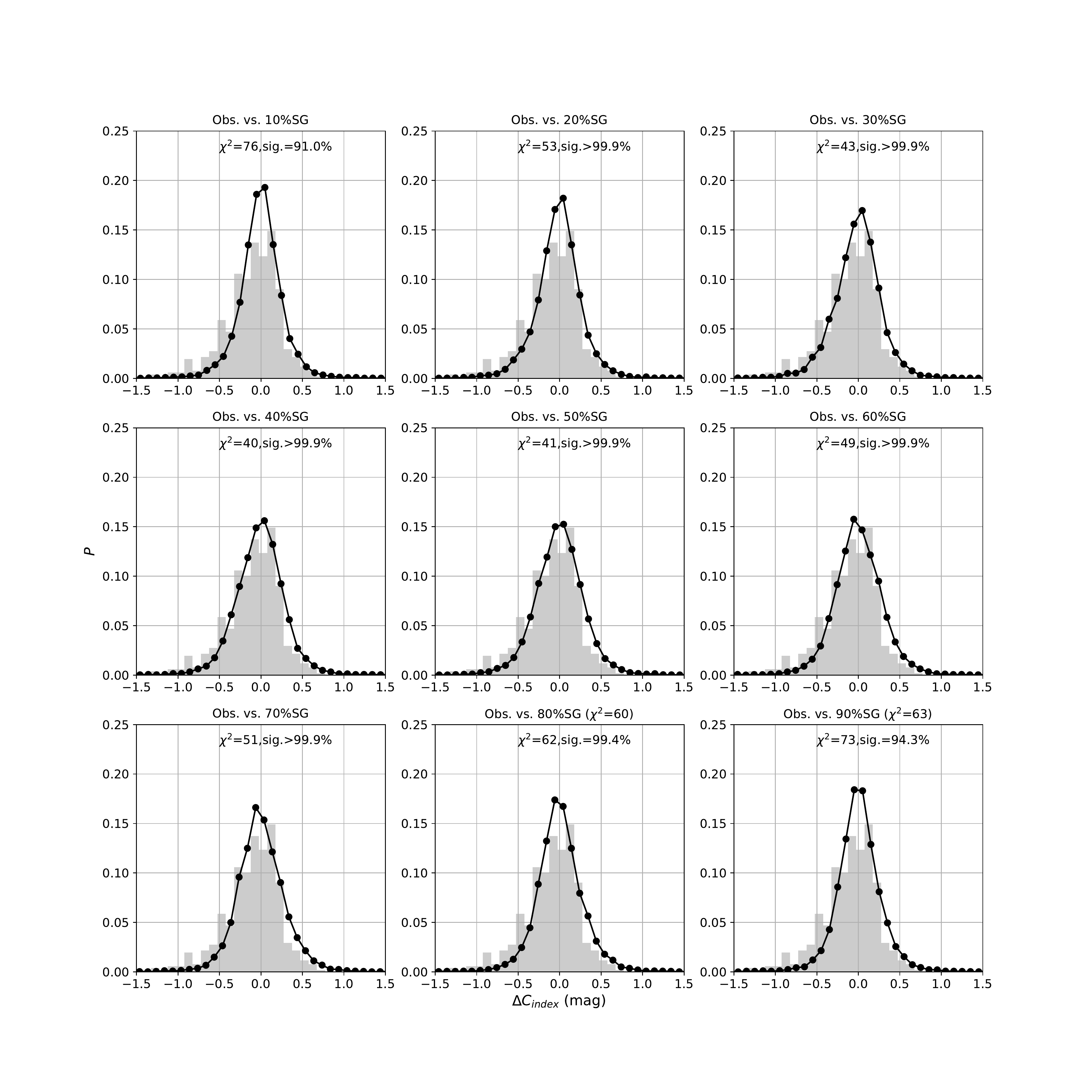}
\caption{As Fig.\ref{F7}, but the chemical spread follows a bimodal distribution 
($\Delta$[N/Fe]=0.0 and 1.0 dex).
}
\label{F8}
\end{figure*}

Bimodal distribution is a very common for chemical spread in clusters with MPs. As an example, 
\cite{Nied17a} find that NGC121 exhibits two distinct RGBs which can be described by a pristine 
population and an enriched population with $\Delta$[N/Fe]=1.1 dex. We employed a 
bimodal distribution which only contains a pristine population and an enriched population with 
$\Delta$[N/Fe]=1.0 dex. We vary the fraction of enriched population stars from 10\% to 90\% and 
calculate the same $\chi^2$ to quantify their similarities to the observation. This result is presented 
in Fig.\ref{F8}. 

In Fig.\ref{F8} we find that with a bimodal chemical distribution, a fraction of 40\% enriched 
stars can make the optimal reproduction. Compared with the optimal reproduction in Fig.\ref{F7}, the 
$\Delta{C_{\rm F275W,F343N,F438W}}$ distributions in Fig.\ref{F8} also exhibit a better fit to the 
observation, with a lower $\chi^2$ (40). The flat and bimodal distributions represent two extreme cases 
of chemical spread, the real distribution of chemical spread should be some shape between these 
two cases. In this work, we therefore conclude that the distribution of chemical spread for the observation 
is more likely a bimodal rather than a flat distribution. The fraction of enriched population stars among 
the GK-type MS of NGC 1978 would range from 40\% (bimodal) to 80\% (flat). 

\begin{figure*}[!ht]
\includegraphics[width=2\columnwidth]{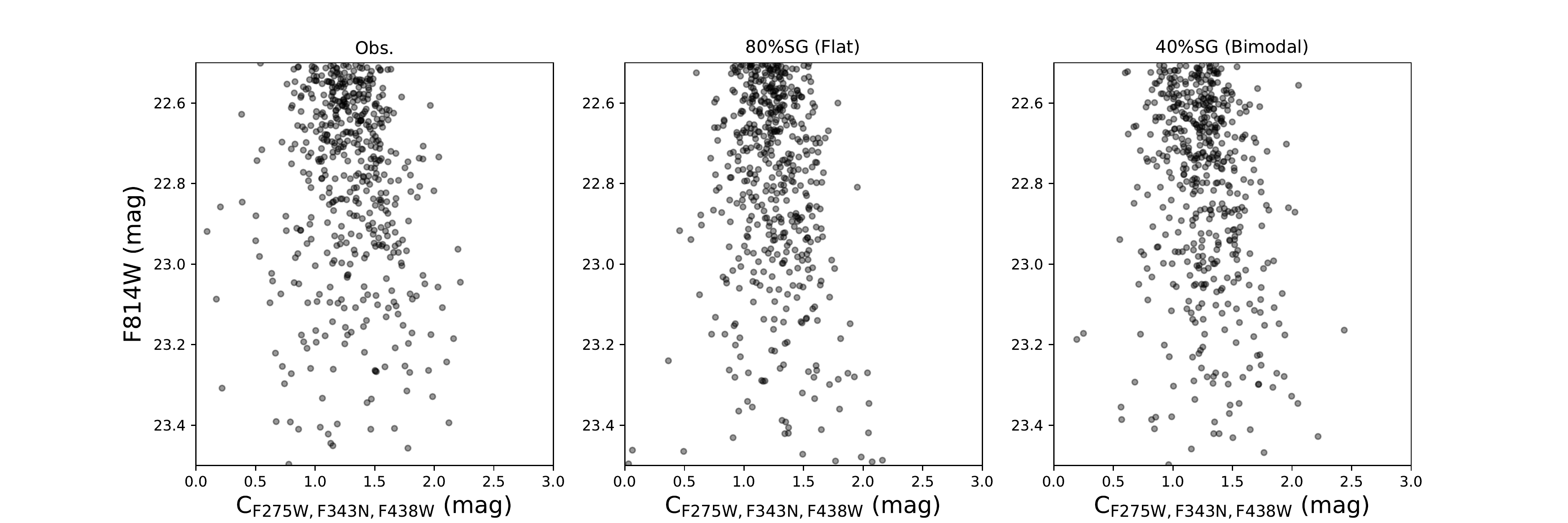}
\caption{$C_{\rm F275W,F343N,F438W}$ vs. F814W diagram of the observation (left), the 
synthetic MPs with 80\% enriched stars (flat chemical spread), and the MPs with 40\% 
enriched stars (bimodal chemical spread).
}
\label{F9}
\end{figure*}

Finally, in Fig.\ref{F9}, we present the observed $C_{\rm F275W,F343N,F438W}$ vs. F814W diagrams 
as well as the `best-fitting' synthetic MPs for two different chemical distributions: a flat distribution 
with 80\% enriched stars and a bimodal distribution with 40\% enriched population stars. 
For the observed $C_{\rm F275W,F343N,F438W}$ vs. F814W diagrams, the field contamination 
has been statistically subtracted based on the selected `reference field' (with completeness and 
area difference corrected). We find that a bimodal distribution with 40\% enriched  
stars exhibits a higher similarity to the observation than the flat distribution model (70\% enriched 
stars). 

\section{Discussion and conclusion}

The very large 
photometric uncertainties prevent us from resolving the stellar populations in detail. 
As shown in fig. \ref{F6}, an internal chemical spread of $\delta$[N/Fe]$\leq$0.6 dex would 
not be very different from the SSP model. As a result, even if these dwarfs are MPs with some 
modest chemical spread, our analysis could not uncover this signature. In this work, to minimize the 
effect of large photometric uncertainties, we have considered many physical and artificial effects, 
including differential reddening, PSF fitting variations and color variations. Some 
of these effects can be partially corrected in our ASs sample, and some have to be 
considered as an additional noise that we cannot reduce. Given that we have considered 
all (major) uncertainties and contaminations, the presence of MPs is the most reasonable 
explanation to the observed color spread of the MS.

\subsection{Radial distribution of multiple populations.}
A recent study has revealed a clear evolution of the spatial mixing between pristine- 
and enriched stars as a function of dynamical stage among different 
GCs \citep{Dale19a}. The dynamical evolution of clusters can readily explain this result with an 
initially more concentrated enriched population. If different population stars indeed cause the 
broadening of the MS of NGC 1978, we may see the difference in their central concentration. 
We employed a poor man's method by dividing the GK-type stars into two subgroups 
with $\Delta{C_{\rm F275W,F343N,F438W}}>$0 and $\Delta{C_{\rm F275W,F343N,F438W}}<$0. 
Based on our model, the sample with negative $\Delta{C_{\rm F275W,F343N,F438W}}$ would  
contain more enriched stars than the sample with positive $\Delta{C_{\rm F275W,F343N,F438W}}$. 
We thus define the sample with $\Delta{C_{\rm F275W,F343N,F438W}}>$0 as the `first generation' (FG), and its counterpart as the SG. Based on \cite{Dale19a}, we should see that the SG is 
more centrally concentrated than the FG because NGC 1978 is much younger than most GCs. 

\begin{figure}[!ht]
\includegraphics[width=\columnwidth]{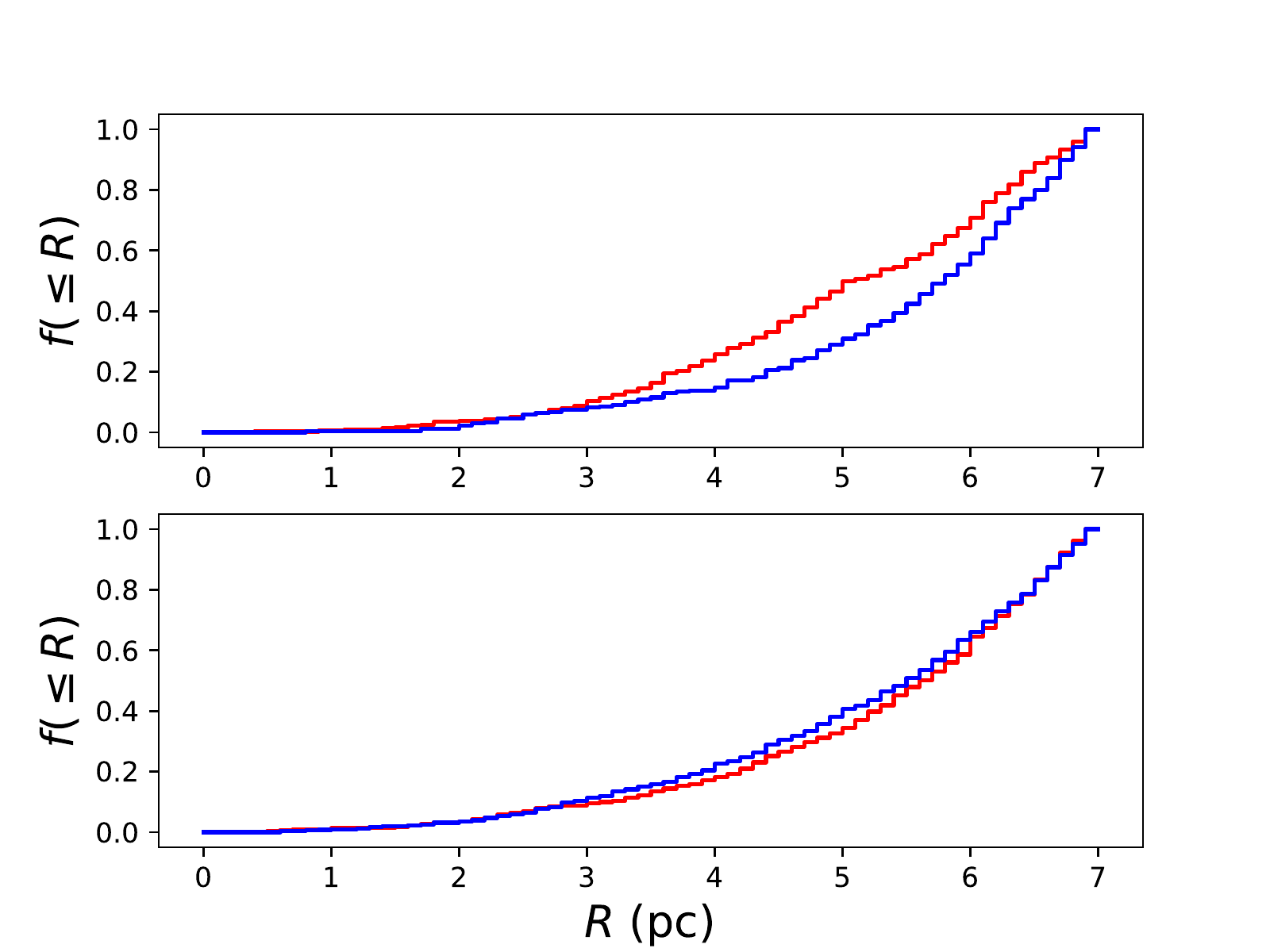}
\caption{Top panel: cumulative curves for stars with positive (blue) and negative (red) $\Delta{C_{\rm F275W,F343N,F438W}}$. Bottom panel: as the top panel, but for ASs. 
}
\label{F10}
\end{figure}

In the top panel of Fig.\ref{F10}, we show the cumulative curves of `FG' and `SG' as functions 
from the cluster center to $R_{\rm hl}$. We find a very significant difference between the `FG' and `SG', 
where the `SG' is indeed more centrally concentrated than the `FG'. We then explore if this 
concentration difference is simply caused by observational artifacts such as blending or 
different sky levels for stars with different $\Delta{C_{\rm F275W,F343N,F438W}}$. This 
is examined through AS tests. In the bottom panel of Fig.\ref{F9}, we did not find any significant 
difference in central concentration of ASs with positive/negative $\Delta{C_{\rm F275W,F343N,F438W}}$. Actually, contrary to the observation, we find that ASs with positive $\Delta{C_{\rm F275W,F343N,F438W}}$ are slightly more centrally concentrated than ASs with negative $\Delta{C_{\rm F275W,F343N,F438W}}$. This is caused by the presence of unresolved binaries. Therefore the fact that the SG would have a higher central concentration than the FG cannot be explained by observational artifacts. The observed different concentrations between the FG and SG are a physical reality. 

\subsection{The fraction of enriched population stars}
Accurate determinations of the fraction of N-rich stars is challenged by photometric 
uncertainties and residual differential reddening. The amount of differential 
reddening associated to each star corresponds to the median reddening of a sample 
of 30 neighbors \citep[see][]{Milo12a}. Hence, our differential-reddening correction 
does not account for very small-scale reddening variations. Because the internal 
spreads of differential reddening and color variations for these nearest stars cannot be 
corrected, and they will contribute to additional broadening of the MS as well. 
That means we may have overestimated the fraction of enriched stars. 
But this overestimation should be minor, as the typical residuals of the 
differential reddening and color variations, could be only roughly their internal spread 
(after correction) divided by the number of nearest stars we used. In summary, 
if we assume a bimodal distribution in chemical spread, the minimum fraction of 
enriched population stars is $\sim$40\%, which is more than twice that found by 
\cite{Milo20a} (15\%). 

We estimate the initial mass of NGC 1978. If we assume that the initial stellar 
mass function (IMF) of NGC 1978 is Kroupa-like, depending on polluters, the mass 
fractions of different polluters varies from 7\% \citep[massive binaries][]{deMi09a} 
to 15\% \citep[intermediate-mass AGB stars][]{Bekk17a}. If assuming a realistic maximum 
star formation efficiency of 30\% \citep[][]{Lada03a}, the initial mass of NGC 1978 
should be at least nine times its current mass. 

We then evaluated the dynamical age of NGC 1978. We calculated the 
brightness profiles of NGC 1978 in F275W and F814W passbands. We then 
fitted these brightness profiles using the Elson-Fall-Freeman (EFF) function \citep{Elso89a}.
The brightness profiles in F275W and F814W are mutually consistent with 
each other. We determined an average half-light radius of $r_{\rm hl}=12.0\pm1.7$ pc 
based on their best fitting EFF functions. Next, we counted the number of stars within the 
range below 1 to 2 mag of the turnoff point in the F814W passband. We assume that these stars  
belong to a Kroupa-like mass function \citep{Krou01a}. By extrapolating the 
whole population down to the minimum stellar mass of 0.08$M_{\odot}$, we estimated the total mass 
of NGC 1978 (assuming that its half-light radius is roughly the same as half-mass radius). Our 
result yields $\log{M/M_{\odot}}=5.28^{+0.04}_{-0.06}$ for NGC 1978, which is 
close to \citep{Baum13a}. Finally, the half-mass relaxation time is calculated as \citep{Spit69a},
{\begin{equation}
t_{\rm rh}=\frac{0.78\; {\rm Gyr}}{\ln{\lambda{N}}}\frac{1\; M_{\odot}}{m}\left(\frac{M}{10^5\; M_{\odot}}\right)^{1/2}\left(\frac{r_{\rm hm}}{1\; {\rm pc}}\right)^{3/2}
\end{equation}}
with $\lambda=0.1$ \citep{Gier94a}. Our result yields $t_{\rm rh}=4.62^{+0.46}_{-0.43}$ Gyr, 
which is more than twice the age of NGC 1978. Therefore NGC 1978 must be a dynamically young 
stellar system. 

However, our results is in contradiction with numerical simulations, which show that FG 
and SG stars will be fully mixed when a cluster has lost 60$-$80 percent of its initial mass \citep{Vesp13a,Miho15a}. 
If the numerical simulations are correct, it may indicate that 
the initial mass of NGC 1978 is not such high. For example, its IMF may be top-heavy 
rather than Kroupa-like. In any case, so far no embedded cluster is known to have such 
a high initial mass in the Magellanic Clouds. As suggested by \cite{Li19b}, searching for 
young clusters with initial masses higher than this mass threshold would allow to 
disentangle among the various scenarios for the formation of MPs.


\subsection{Summary}
Our results indicate that MPs are present among the GK-type MS stars of NGC 
1978. This is expected since NGC 1978 is already confirmed to harbor MPs among 
its red-giant stars \citep{Milo20a}. In our Fig.\ref{F1}, we also see a clear 
$C_{F275W,F343N,F438W}$ broadening of its RGB. This result confirms that the MPs 
in NGC 1978, like most GGCs, are a global feature for stars at different stellar evolutionary 
stages. Since the unevolved GK-type dwarfs are not hot enough to trigger the CNO-chain, 
their chemical anomaly must be produced through other polluters, i.e., massive stars. 


In summary, after our study of NGC 419, this work is the second attempt of searching 
for MPs among GK-type dwarfs in clusters with ages around $\sim$2 Gyr. We summarize 
our results as follows: 

\begin{itemize}
\item[{\bf a.}] The GK-type MS dwarfs in NGC 1978 exhibit a large $C_{F275W,F343N,F438W}$ spread, which 
can not be entirely reproduced by observational uncertainties. This fact demonstrates that NGC 1978 hosts MPs among MS stars. 
\item[{\bf b.}] To reproduce the observed $\Delta{C_{\rm F275W,F343N,F438W}}$ distribution, we 
require an internal chemical spread of $\delta$[N/Fe]=1.0 dex. 
\item[{\bf c.}] By assuming a flat distribution of the C, N, O stellar contents, we find that a fraction of 60\%–80\% SG stars is required to properly reproduce the observations. A bimodal distribution of light elements composed of 40\% of 2G stars, would provide a better match with the observations than that provided by flat elemental spread.  
\item[{\bf d.}] Stars with negative $\Delta{C_{\rm F275W,F343N,F438W}}$ are more centrally concentrated 
than stars with positive $\Delta{C_{\rm F275W,F343N,F438W}}$. 
\end{itemize}

We conclude that NGC 1978 harbors a chemical spread of up to $\delta$[N/Fe]=1.0 dex (with $\delta$[C/Fe]=
$\delta$[O/Fe]=$-$0.63 dex) among its GK-type MS stars. The fraction of enriched stars is at least 40\%. 
The pristine and enriched stars are different in spatial distribution, with the enriched stars more 
centrally concentrated than the pristine population stars, which is similar to what is seen in most GCs. We suggest 
that like for most GGCs, MPs are a global feature for stars at different evolutionary stages in extragalactic star clusters.

\acknowledgements 
{C. L. and B. T. acknowledge support from the one-hundred-talent project of Sun Yat-sen 
University. C. L. and Y. W. were supported by the National Natural Science Foundation 
of China through grants 11803048 and 12037090. B. T. gratefully acknowledges support from the National 
Natural Science Foundation of China through grant No. U1931102. 
This work has received funding from the European Research Council (ERC) under the European Union's Horizon 2020 research innovation programme (Grant Agreement ERC-StG 2016, No 716082 `GALFOR', PI: Milone,  \url{http://progetti.dfa.unipd.it/GALFOR}); from MIUR through the FARE project R164RM93XW (SEMPLICE, PI Milone) and through the PRIN program 2017Z2HSMF (PI Bedin). J. H. was supported by Basic Science Research Program through the National Research Foundation of Korea (NRF) funded by the Ministry of Education (No. 2020R1I1A1A01051827). Y. Y. gratefully acknowledges financial support from the China Scholarship Council (grant 201906010218) and 
the National Natural Science Foundation of China through grants 11633005. Y. W. was 
supported by the Special Research Assistant Foundation Project of Chinese Academy of 
Sciences.}

\end{document}